\documentclass[conference]{IEEEtran}

\usepackage{amsfonts}
\usepackage{amsmath}
\usepackage{amsthm}
\usepackage{color}
\usepackage{graphicx} 
\usepackage{multirow}
\usepackage{float}
\usepackage{nicefrac}
\usepackage{verbatim}
\usepackage{subfigure}
\usepackage{balance}
\usepackage{diagbox}
\usepackage{cite}
\usepackage{verbatim}
\usepackage{subfigure}
\usepackage{balance}
\usepackage{diagbox}
\usepackage{enumitem}
\usepackage{diagbox}

\setlist[itemize]{leftmargin=*}
\setlist[enumerate]{leftmargin=*}
\begin{document}

\title{Forecasting Busy-Hour Downlink Traffic in Cellular Networks}

\author{Andrea Pimpinella$^{\ast}$, Federico Di Giusto$^{\ast}$, Alessandro E. C. Redondi$^{\ast}$, Luisa Venturini$^{\dagger}$, Andrea Pavon$^{\dagger}$\\ 
$^{\ast}$Dip. Elettronica, Informazione e Bioingegneria, Politecnico di Milano\\
$^{\dagger}$Vodafone Group, Network Engineering and Delivery \\
Email: {name.surname}@polimi.it or {name.surname}@vodafone.com
}

\maketitle

\newcommand{\bkref}[1] {(\ref{#1})}

\begin{abstract}
The dramatic growth in cellular traffic volume requires cellular network operators to develop strategies to carefully dimension and manage the available network resources. Forecasting traffic volumes is a fundamental building block for any proactive management strategy and is therefore of great interest in such a context. Differently from what found in the literature, where network traffic is generally predicted in the short-term, in this work we tackle the problem of forecasting \textit{busy hour} traffic, i.e., the time series of observed daily maxima traffic volumes. We tackle specifically forecasting in the long term (one, two months ahead) and we compare different approaches for the task at hand, considering different forecasting algorithms as well as relying or not on a cluster-based approach which first groups network cells with similar busy hour traffic profiles and then fits per-cluster forecasting models to predict the traffic loads. Results on a real cellular network dataset show that busy hour traffic can be forecasted with errors below 10\% for look-ahead periods up to 2 months in the future. Moreover, when clusters are available, we improve forecasting accuracy up to 8\% and 5\% for look-ahead of 1 and 2 months, respectively. 
\end{abstract}


\section{Introduction}\label{sec:intro}
In recent years cellular networks have witnessed a dramatic growth of traffic volume, which is estimated to reach 226 EB per month in 2026. This is mainly due to the prosperity of mobile subscriptions (which are expected to overcome 9 billions in 2026), the improved device capabilities and an increase of data-intensive contents\cite{ericsson}. To cope with this situation, Mobile Network Operators (MNOs) are gradually implementing efficient resource management strategies. However, the increased complexity of mobile network dynamics has raised MNOs awareness that the efficiency of network planning and dimensioning interventions depends on how well mobile data traffic can be analysed, understood and modeled especially in urban contexts.

A way MNOs have to implement proactive resources allocation strategies is to design traffic load \textit{forecasting} algorithms, with look-ahead horizons (i.e., how far in the future traffic is forecasted) which go from several hours to several months\cite{trinh2018mobile, zhang2018long, le2018applying, mahdy2020clustering}. On the one hand, the development of powerful hardware capable of dealing with the massive amount of data coming from cellular networks have enabled the optimisation of traffic forecasting models. On the other hand, making the network aware of traffic demands and capable of pro-actively responding to traffic dynamics is yet a complex task. 

At the same time, it has been shown \cite{furno2016tale, shi2018discovering, xu2016big} that there is a strong relationship between users mobile communication activity and the characteristics of the underlying urban area where such activity takes place. The goal of grouping network sites according to the spatial and temporal dynamics of the served traffic is typically addressed through the design of \textit{clustering} algorithms. From a spatial domain perspective, the knowledge of a clustering configuration can be crucial for MNOs considering that traffic loads are often characterised by geographic shifts between precise urban areas during a day and during weekday-to-weekend transitions \cite{furno2016tale}. Considering the time domain, the similarity of mobile traffic dynamics within a cluster can be exploited to improve the accuracy of traffic modeling techniques and thus discover regularities of traffic demand peaks.



In this paper we focus on forecasting traffic volumes starting from a real dataset collected from the Vodafone mobile network of a middle-sized city in northern Italy. Differently from related works, we consider as forecasting target the time series of downlink traffic at the \textit{busy} \textit{hour}, i.e., the time series of daily maxima traffic loads. We compare different forecasting algorithms (time series decomposition, ARIMA and Long Short Term Memory networks), also considering the possibility of clustering network cells prior to computing forecasts. Results show that busy hour traffic can be forecasted with errors below 10\% for look-ahead periods up to 2 months in the future. Moreover, when clusters are available, we improve forecasting accuracy up to 8\% and 5\% for look-ahead of 1 and 2 months, respectively.


The rest of this paper is structured as it follows: Section \ref{sec:sota} summarises related works about clustering and traffic forecasting in cellular networks, while Section \ref{sec:methodology} describes the dataset available for this work and the approach we follow to both sites clustering and forecasting. Section \ref{sec:results} comments the designed clustering configuration as well as the results obtained from forecasting busy hour downlink traffic, when either clusters are unknown or when the produced clustering configuration is available. Finally, Section \ref{sec:conclusions} concludes the paper.

\section{Related Works}\label{sec:sota}
Several works in the literature have analysed cellular networks traffic traces to understand and model traffic patterns, especially in urban environments \cite{furno2016tale, shi2018discovering} . 
In \cite{furno2016tale} a heterogeneous dataset containing mobile traffic of ten international cities is investigated with the goal of i) summarising the mobile traffic activity in each area and ii) grouping similar area signatures into a limited representative set. In particular, leveraging the $k$-means algorithm, the authors recognise five different area types, namely residential, office, transportation, touristic and leisure. With similar aims, in \cite{shi2018discovering} authors consider a large-scale cellular network in Shangai (9K base stations for a total of 3M users) and propose a model to cluster traffic profiles with the goal of assisting MNOs in network planning operations. Again, results show that 5 clusters are recognised which the authors label as residential, transport, office, entertainment and comprehensive (the latter grouping those base stations not associated to the other clusters). 
In many cases, clustering models are used to improve the performance of traffic load forecasting algorithms. In fact, a common approach to traffic forecasting is first to cluster network sites with respect to weekly or monthly traffic distribution (in space or time) and secondly fit a prediction model within each cluster \cite{shi2018discovering, mahdy2020clustering, le2018applying}. Authors in \cite{shi2018discovering} show that clusters knowledge improve the average forecasting accuracy up to 20\% with respect to the case when no clustering is available. This is also shown in \cite{mahdy2020clustering}, where authors leverage a clustering algorithm to improve the performance of several forecasting methods. After training the selected models i) on the cumulative traffic load and ii) on each cluster's cumulative load, results show that clusters knowledge benefits the training process and improve the average accuracy by 25\%. This comparison is also performed in \cite{le2018applying}, with similar conclusions: forecasting performance improve when an ad-hoc model is developed per each recognised cluster.

Differently from the aforementioned works, we consider as target the busy hour traffic, with forecasting horizons larger than 1 month. This is of great interest for MNOs, which often focus on the implementation of dimensioning and resources allocation strategies in a pro-active rather than reactive fashion.



\section{Methodology}\label{sec:methodology}
This work considers a dataset coming from Vodafone, one of the major European mobile operators. The dataset contains radio access network measurements relative to the period $T$=\{06/01/2020, 31/07/2020\} and referring to about 1.500 eNodeBs of the Vodafone LTE network in the city of Milan, Italy. In particular, for each eNodeB, we exploit the availability of several network Key Performance Indicators (KPIs) (e.g., cumulative downloaded and uploaded data volumes, n. of handovers, etc.) and the congestion level at the cell site (e.g., n. of active connections, n. of PRBs used, etc.) in the form of hourly sampled time series. In this work, we consider the cumulative traffic downloaded at each cell site with a focus on the traffic served during the \textit{busy hour}, as detailed in the next Section.

\subsection{Definition of Busy Hour}\label{subsec:busyhour}
Let $v^e(t)$ be the time series representing the hourly traffic volume downloaded from the $e$-th eNodeB. The total hourly downlink traffic can be written as:
\begin{align}
& \mathbf{v}(t) = \sum_{e} \hspace{2mm} {v^e}(t)
\end{align}
From this time series we build a new time series $\mathbf{v_{b}}(d)$ containing only daily busy hours samples, by picking, for each day $d$, the traffic sample having the maximum value\footnote{Beside the considered traffic-based definition of the busy hour, this can also be defined as the hour of the day which maximizes the number of connected users. In this work we will not consider this second option, as the downlink traffic sampled at the corresponding busy hours is in our case always lower in volume than the one sampled in the traffic-based busy hours.}.


\begin{figure*}[t!]
\centering
\includegraphics[width=1.9\columnwidth]{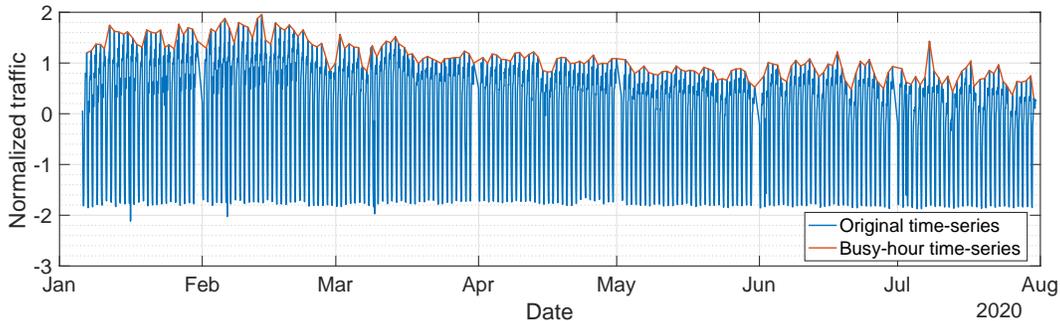} 
\caption{Normalised cumulative downlink traffic time series (blue) and the corresponding busy hour time series (red) for all the eNodeBs of the network, on the whole period $\mathcal{T}$.}
\label{fig:dl_bh}
\end{figure*}

Figure \ref{fig:dl_bh} depicts the evolution of $\mathbf{v_{b}}(d)$ in the period of interest, where traffic volumes have been normalised for privacy reasons. As one can see, the time series is characterized by a slightly increasing trend from January 2020 to March 2020 while the trend is decreasing from March 2020 onwards. Also, a strong weekly seasonality can be observed before $09/03/2020$ and after $01/06/2020$, while traffic patterns appear unregular between these two dates\footnote{In Italy, full lockdown has been established between $09/03/2020$ and $01/06/2020$ due to COVID-19 pandemic.}. 

\subsection{Forecasting Methods}\label{subsec:forecasting}
Considering the interest of MNOs in predicting when the next peak of traffic demand will occur, in this work we focus on the forecast of $\mathbf{v_{b}}$. We compare the performance of three different forecasting methods:
\begin{itemize}
	\item \textit{Additive Time Series Decomposition (AD)}: we assume $\mathbf{v_{b}}$ to equal the sum of three independent components:
		\begin{equation}
			\mathbf{v_{b}} =S + T + E
			\label{eq:additive_decmp}
		\end{equation}
		where $S$, $T$ and $E$ are the \textit{Seasonal}, \textit{Trend} and \textit{Noise} components, respectively. While $E$ is by nature not predictable, $S$ and $T$ can be independently forecasted through a traditional \textit{na\"ive} decomposition model and the concatenation of a \textit{STL}\footnote{The acronym stands for Seasonal and Trend decomposition using Loess, a model used for time series decomposition.} decomposition model and a non-seasonal \textit{ARIMA}\cite{hyndman2018forecasting}, respectively. Note that this procedure requires in input the expected \textit{periodicity} $\tau$ of the time series, which can be retrieved observing the time plot of the data.
		
	\item \textit{Seasonal ARIMA (SA)}: a well-known approach for stationary seasonal time series forecasting is to model the forecast target as the combination of two independent ARIMA models, one for the non-seasonal part of the time series and the other one for the seasonal part. The combined model is denoted as ARIMA$(p,d,q) \times (P,D,Q)_S$ where $(p,d,q)$ are the AR, Integration and MA order of the non-seasonal part while $(P,D,Q)$ are the corresponding seasonal versions, being $S$ the period of the repeating seasonal pattern  \cite{siami2018comparison}. A crucial step for modelling a time series through ARIMA is to estimate the different orders which characterise the model: we will detail this aspect later in Section \ref{subsec:tuning}.    
	\item \textit{Long Short-Term Memory (LSTM)}: another way to forecast time series is to assume a more complex, non linear relationship between data samples and learn it by means of Recurrent Neural Networks (RNNs)\cite{schmidhuber2015deep, siami2018comparison}. LSTM structures are characterized by the concatenation of $h$ hidden layers, where each layer is formed by $n_h$ \textit{cells}. 
A common approach is to concatenate at the end of the LSTM structure $k$ fully connected hidden layers (with $m_k$ neurons each) and a final layer (with $O$ output units) to output the forecasted samples\cite{siami2018comparison}. Note that the size $O$ of the output layer represents the look-ahead forecast horizon, i.e., how far in the future the network is able to forecast. Details on the tuning of these hyper-parameters are given in Section \ref{subsec:tuning}.
\end{itemize}

\subsection{Clustering Procedure}\label{subsec:clustering}
The forecasting methodologies presented before can be applied directly on the aggregate busy hour time  series $\mathbf{v_{b}}$ to perform prediction. An alternative approach is to cluster together eNodeBs having similar busy-hour traffic behaviour, perform forecasts for each cluster and finally sum the obtained prediction together. We refer to the former approach as Cluster-Unaware (CU), while the latter is named Cluster-Aware (CA).

\begin{figure}[t]
\centering
\includegraphics[width=0.8\columnwidth]{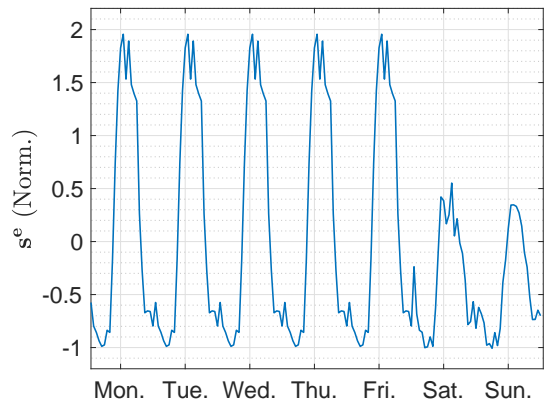} 
\caption{Normalised MWS for one eNodeB of example: working day MDS is repeated from Monday to Friday while Saturday and Sunday have different MDS.}
\label{fig:mws}
\end{figure}

The Cluster-Aware approach relies on $k$-means clustering, grouping together eNodeBs based on the Euclidean distance among their Median Weekly Signatures (MWS). In details, the MDS of an eNodeB is computed as it follows:
\begin{itemize}
	\item First, we compute the Median Daily Signature (MDS) for work days (Mondays to Fridays), Saturdays and Sundays. In a nutshell, the MDS for week days is computed hourly by taking the median of the downlink traffic observed at each hour for each day in the dataset. The same is repeated independently for Saturdays and Sundays. 
	\item Then, we compute the MWS by stacking five copies of the work days MDS plus the MDS for Saturdays and Sundays. As an example, we plot in Figure \ref{fig:mws} the (normalised) MWS of one eNodeB taken as example: as one can see, the traffic pattern is the same from Monday to Friday whereas the pattern changes during the week end. 
\end{itemize}

Repeating this process for each eNodeB in the network allows to perform $k$-means clustering with the MWS as input. Note that each MWS is normalised before being used, as our interest is to group eNodeBs regardless of the amplitude of the observed download traffic. Also, considering that the first COVID-19 infection case was registered in Italy on 15/02/2020, to avoid biasing errors due to the extraordinary traffic behaviour that has been observed afterwards, we compute the MWSs considering only the 4-weeks season observed in January 2020. 

\begin{figure}
\begin{subfigure}
	\centering
	\includegraphics[width =0.95\columnwidth]{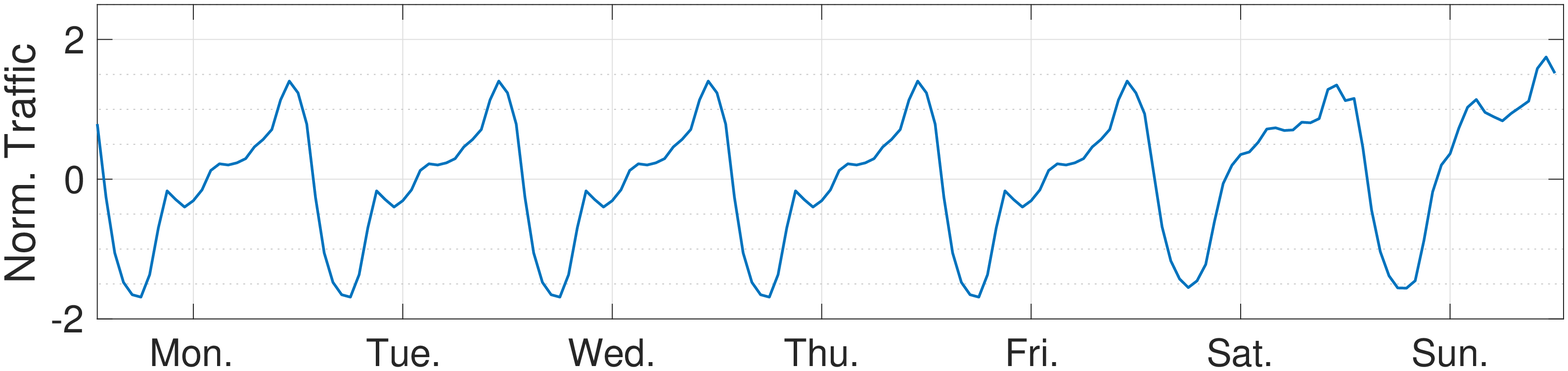}
	\caption{Centroid representative of $\mathcal{R}_1$ cluster.}
	\label{fig:Res1_jan}
\end{subfigure}
\begin{subfigure}
	\centering
	\includegraphics[width =0.95\columnwidth]{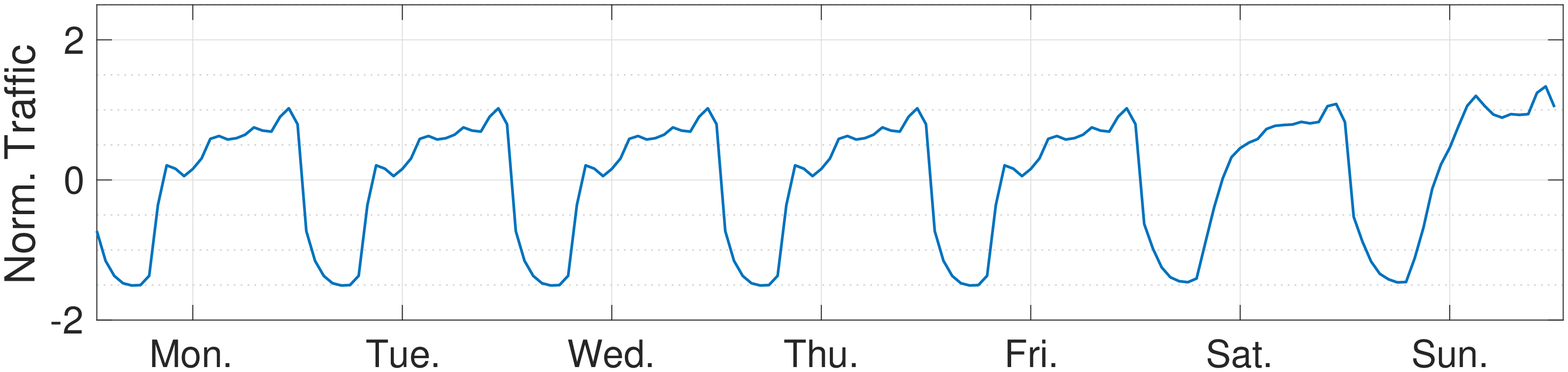}
	\caption{Centroid representative of $\mathcal{R}_2$ cluster.}
	\label{fig:Res2_jan}
\end{subfigure}
\begin{subfigure}
	\centering
	\includegraphics[width =0.95\columnwidth]{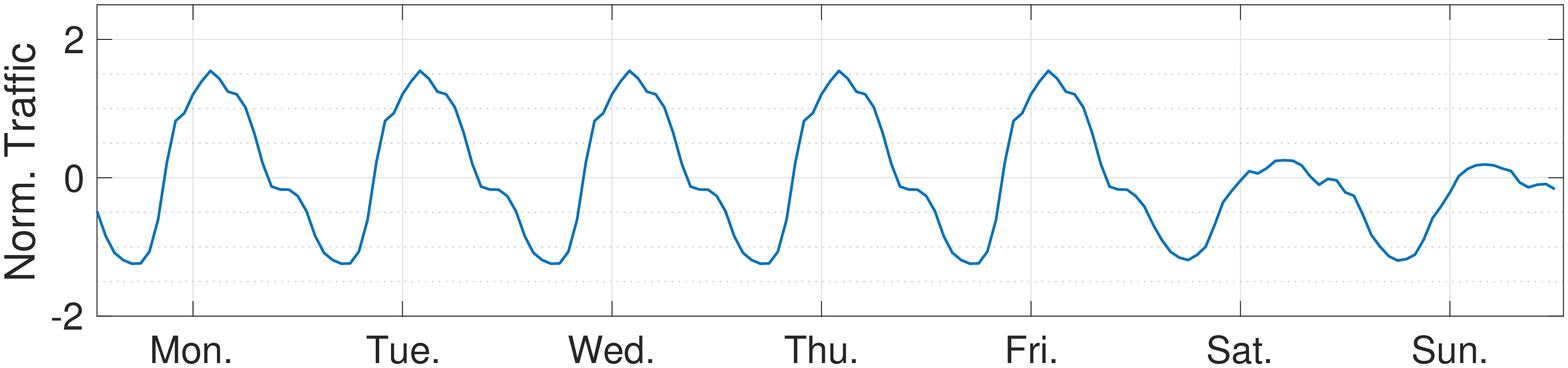}
	\caption{Centroid representative of $\mathcal{B}$ cluster.}
	\label{fig:Bus_jan}
\end{subfigure}
\begin{subfigure}
	\centering
	\includegraphics[width =0.95\columnwidth]{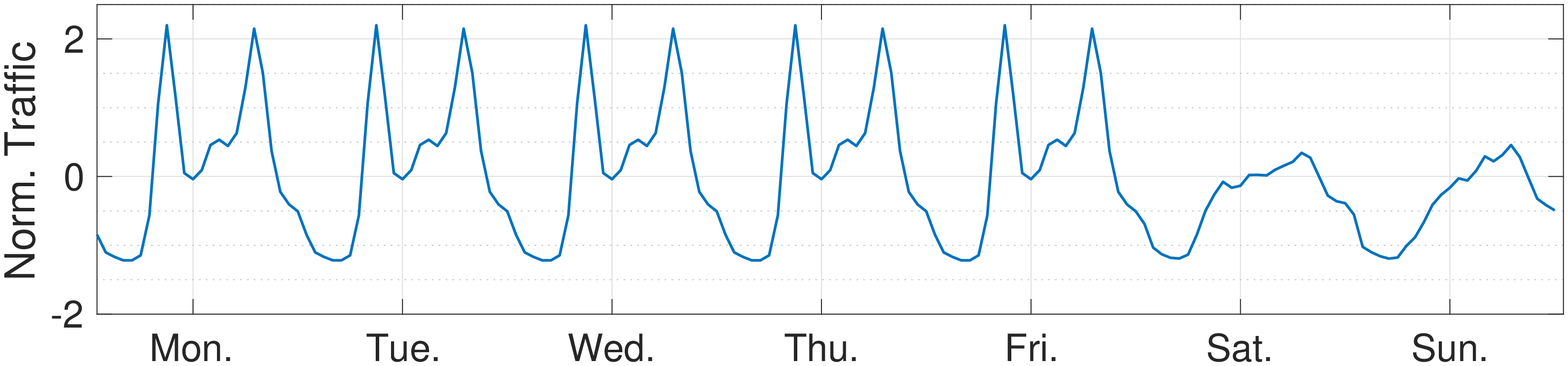}
	\caption{Centroid representative of $\mathcal{T}$ cluster.}
	\label{fig:transp_jan}
\end{subfigure}
\begin{subfigure}
	\centering
	\includegraphics[width =0.95\columnwidth]{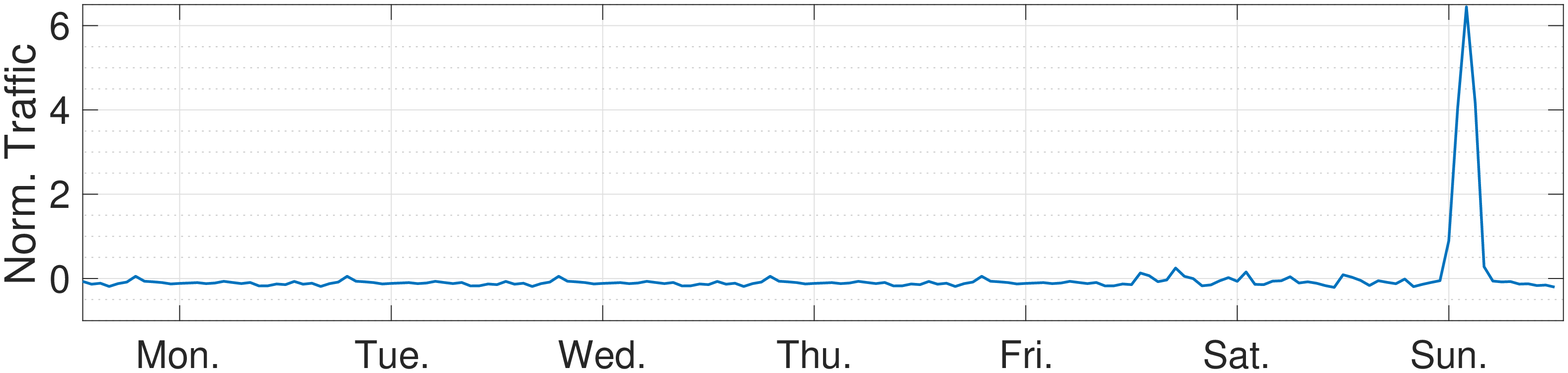}
	\caption{Centroid representative of $\mathcal{U}$ cluster.}
	\label{fig:Unc_jan}
\end{subfigure}
\end{figure}

There is no a-priori knowledge about the number of clusters $k$ to use: therefore, we take a data driven approach which consists in a grid search over the set of candidate values $k=[1, \dots, 10]$. For each candidate value, we perform clustering and compute for each eNodeB the corresponding \textit{Silhouette} coefficient \cite{berkhin2006survey}. The higher the coefficient of a signature, the better it matches its own cluster, such that a clustering configuration is appropriate if most of the signatures have a high Silhouette value. At the end of this process, we set $k=5$, which is also the most likely choice for urban scenarios \cite{furno2016tale,shi2018discovering}. We plot in Figures \ref{fig:Res1_jan}-\ref{fig:Unc_jan} the centroids of the recognised clusters, i.e., the representatives of the MWSs of the eNodebs of each cluster. The MWS shown in Figure \ref{fig:Res1_jan} is characterized by two traffic peaks during Week Days, one weaker peak in the morning between 8:00 a.m. and 9:00 a.m and a second sharper peak in the evening between 9:00 p.m. and 10:00 p.m., whereas during Saturday and Sunday only the evening peak survives. In fact, this cluster (which we label as $\mathcal{R}_1$) is associated to \textit{residential} areas of the considered city, where traffic demand is uniform in time but lower loads are observed during working hours. Similar observations can be made if we look at the centroid represented in Figure \ref{fig:Res2_jan} ($\mathcal{R}_2$), a secondary residential cluster where morning and evening peaks are less evident during Week Days whereas Saturday and Sunday profiles are close to those observed in \ref{fig:Res1_jan}. A different behaviour is instead shown in Figure \ref{fig:Bus_jan}, where traffic is heavier during working hours in Week Days while it is consistently lower in volume otherwise. 
We therefore associate this cluster ($\mathcal{B}$) to the \textit{business} areas with a typical European working time during Week Days from 9:00 a.m. to 6:00 p.m. A further different trace is depicted in Figure \ref{fig:transp_jan}, which represents the cluster associated to \textit{transport} hubs and transportation network in the city ($\mathcal{T}$). In fact, the eNodeBs of this cluster are characterised by high traffic demands during commuting rush hours in the Week Days (i.e., at 8:00 a.m. and 6:00 p.m.) while very weak traffic activity is observed in the week end. 
\begin{table}[t]
\centering
\caption{Distribution of clustered eNodeBs and per-cluster cumulative served traffic demand.}
\begin{tabular}{|c|c|c|}
\hline
\textbf{Type}   & \textbf{Size (\%)} & \textbf{Served Traffic (\%)} \\ \hline
$\mathcal{R}_1$ & 37.1               & 48.6                         \\ \hline
$\mathcal{R}_2$ & 30.2               & 29.1                         \\ \hline
$\mathcal{B}$        & 20.6               & 13.9                         \\ \hline
$\mathcal{T}$      & 7.8                & 8.3                          \\ \hline
$\mathcal{U}$    & 4.4                & 0.1                          \\ \hline
\end{tabular}
\label{table:cluster_details}
\end{table}
We summarise in Table \ref{table:cluster_details} the fraction of eNodeBs of the network grouped in each cluster and the corresponding served traffic demand. As one can see, more than 65\% of the eNodeBs are deployed in residential areas and generate more than 75\% of the mobile traffic. Next most serving cluster is those active in business areas, which groups 20\% of the eNodeBs for a total served demand of 14\%, while Transport cluster contains sligthly less than 8\% of the eNodeBs which overall serve sligthly more than 8\% of the traffic demand. Finally, we observe that a small group of eNodeBs (4.4\%) is represented by a MWS which resembles a noisy process (depicted in Figure \ref{fig:Unc_jan} and referred to as $\mathcal{U}$) and is associated with a negligible portion (0.1 \%) of the overall generated traffic.

\section{Forecasting Results}\label{sec:results}
In this Section, we comment on the tuning of AD, SA and LSTM and on the corresponding performance when they are applied to forecast $\mathbf{v_{b}}$. On the one hand, each model will have a look-back window of 2-weeks (i.e., will take in input 14 past daily samples of $\mathbf{v_{b}}$). This choice comes after a grid search tuning and represents a good compromise between size of the training data and freshness of the information. On the other hand, we set the models basic look-ahead horizon to 7 days, such that if predictions further than 1 week ahead are targeted the models will recursively use forecasted samples as input for further iterations (i.e., to feed the look back window of 14 samples when ground truth data are not available). Note that both training and test data are first transformed through a box-cox normalisation before feeding the forecasting models, as it is commonly done in the field of time series forecasting for variance stabilisation purposes \cite{hyndman2018forecasting}. 
\subsubsection{Model Tuning}\label{subsec:tuning}
The selected models require the tuning of several parameters. For what regards AD, considering that $\mathbf{v_{b}}$ is characterised by a weekly seasonality, we set the periodicity $\tau$ equal to 7 days. For what regards SA, after the inspection of the Auto-Correlation and the Partial Auto-Correlation Functions (ACF and PACF respectively) of $\mathbf{v_{b}}$, we select a model of type $(1,1,0) \times (1,1,0)_7$. This is because: i) the cascade of two differencing operations of orders $d=1$ and $D=1$ returns a stationary time series, ii) the seasonality of $\mathbf{v_{b}}$ has a period of $S=7$ days, iii) the presence of a sharp correlation peak at lag=1 in the PACF suggests the autoregressive orders are $p=1$ and $P=1$ and iv) the presence of several modest correlation peaks in the ACF suggests that moving average processes should be excluded from the model (i.e., $q=0$ and $Q=0$).
Finally, considering the LSTM model, we manually design a encoder-decoder based network structure composed by: i) a couple of encoder-decoder layers of $n_1 = n_2 = 200$ cells; ii) a bi-dimensional fully connected layer characterised by $k=7$ parallel hidden layers of $m_1=\dots=m_7=100$ neurons each; iii) an output layer of size $O=7$. Cells and neurons of the network are activated through ReLu activation functions, while MSE is chosen as loss metric. 
\subsubsection{Performance}\label{subsec:performance}
\begin{figure}[t]
\begin{subfigure}
	\centering
	\includegraphics[width =.48\columnwidth]{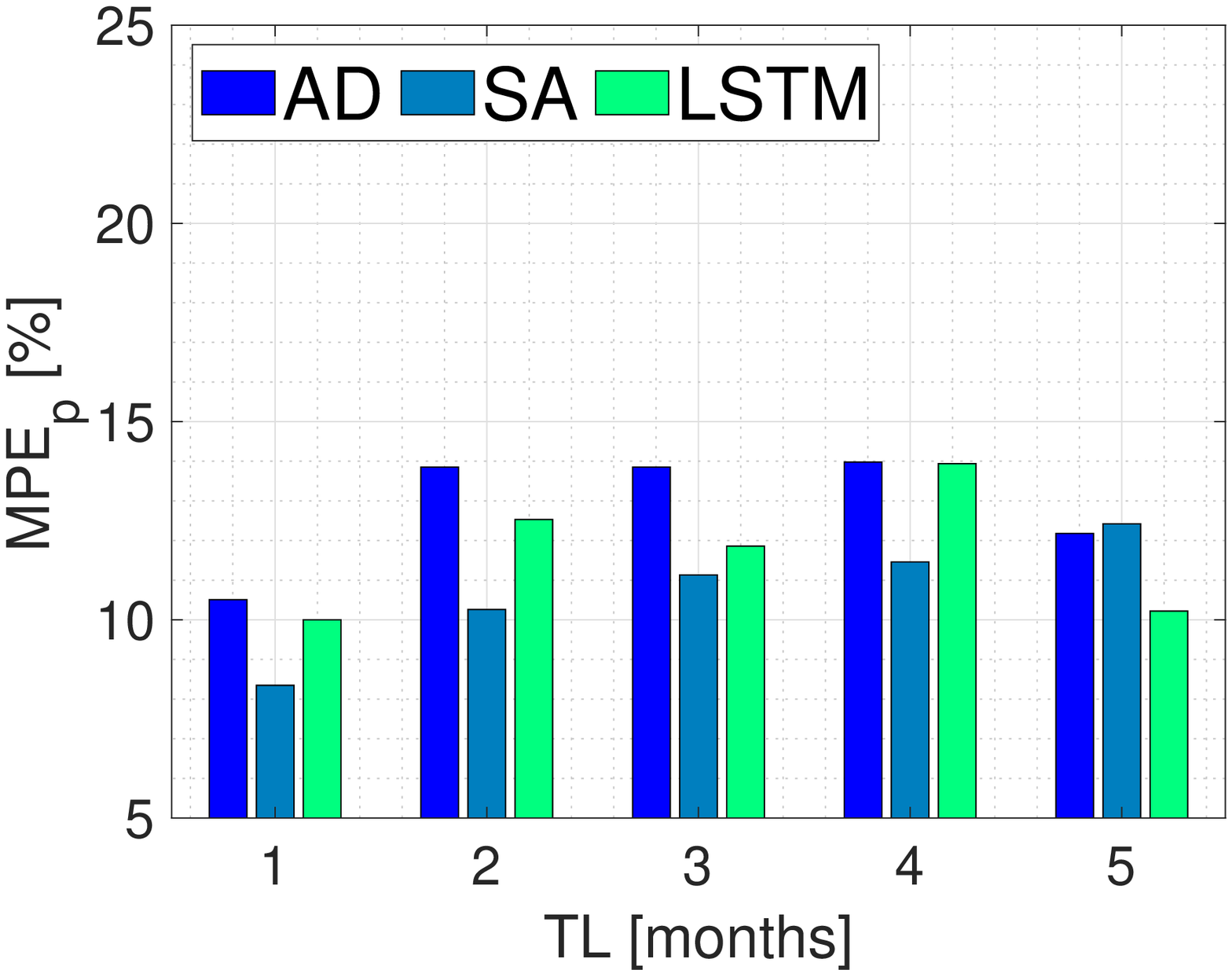}
\end{subfigure}
\begin{subfigure}
	\centering
	\includegraphics[width =.48\columnwidth]{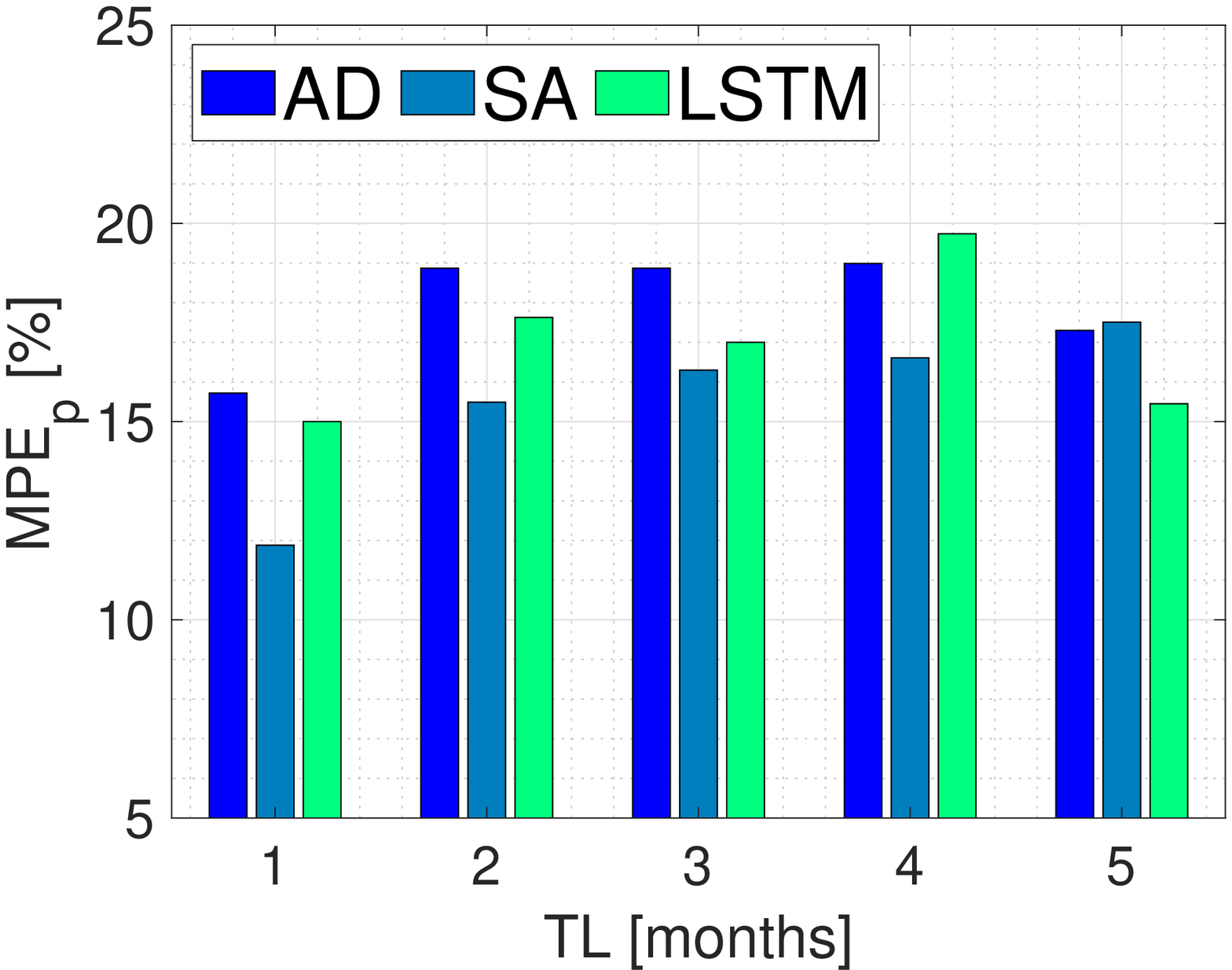}
\end{subfigure}
\caption{MPE, CA: TL vs LA.}
\label{fig:mpe_ca}
\end{figure}

To compare the forecasting performance of the selected methods, we retain the period  $T\textsubscript{train}=\{06/01/2020, 31/05/2020\}$ for training, while the period $T\textsubscript{test}=\{01/06/2020, 31/07/2020\}$ is left aside for testing. The performance are evaluated in terms of Mean Absolute Percentage Error (MAPE) and Mean Percentage Error on Peak (MPE\textsubscript{P}), which are defined as it follows:
\begin{align*}
              &{\rm MAPE} = \frac{100\%}{n}\sum_{t \in T\textsubscript{test}}\left |\frac{v_{b}(t) - \hat{v_{b}}(t)}{v_{b}(t)}\right| \\
              &{\rm MPE\textsubscript{P}} = 100\% \frac{\max\limits_{t \in T\textsubscript{test}} v_{b}(t) - \max\limits_{t \in T\textsubscript{test}} \hat{v_{b}}(t)}{\max\limits_{t \in T\textsubscript{test}} v_{b}(t)}
        \label{eq:metrics}
 \end{align*}
where $\hat{\mathbf{v_{b}}}$ denotes the forecasted time series. Note that MPE\textsubscript{P} measures the capability of a model to predict the maximum value of the forecast target, which is a crucial information when the model is used by a MNO for dimensioning purposes. To study the impact of both the training period length (which we refer to as $TL$) and the look-ahead horizon (namely, $LA$) on the forecasting performance, we compare the results varying $TL$ in [1, 2, $\dots$, 5] and $LA$ in [1, 2] months. 
\begin{figure}[t!]
\begin{subfigure}
	\centering
	\includegraphics[width =.48\columnwidth]{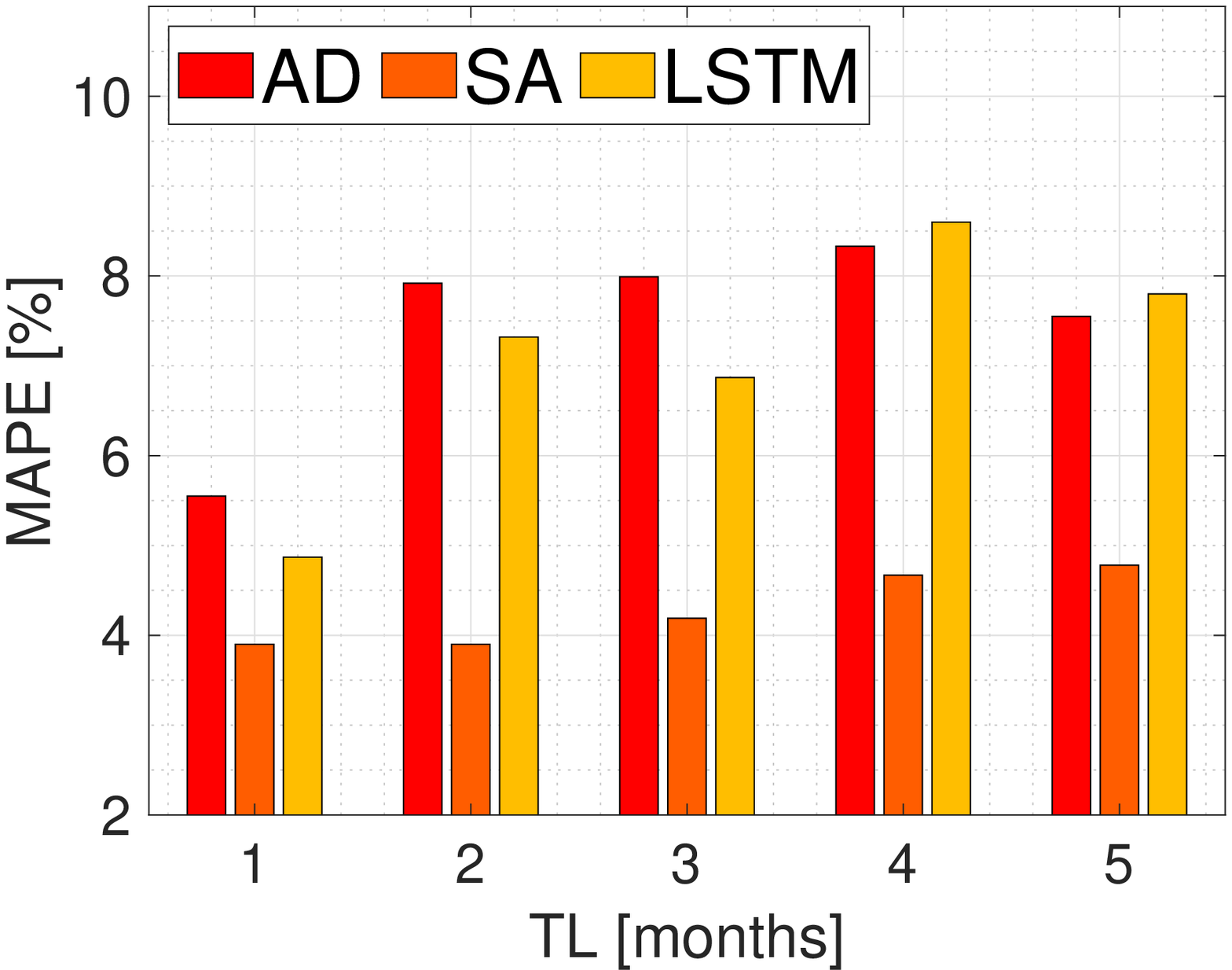}
\end{subfigure}
\begin{subfigure}
	\centering
	\includegraphics[width =.48\columnwidth]{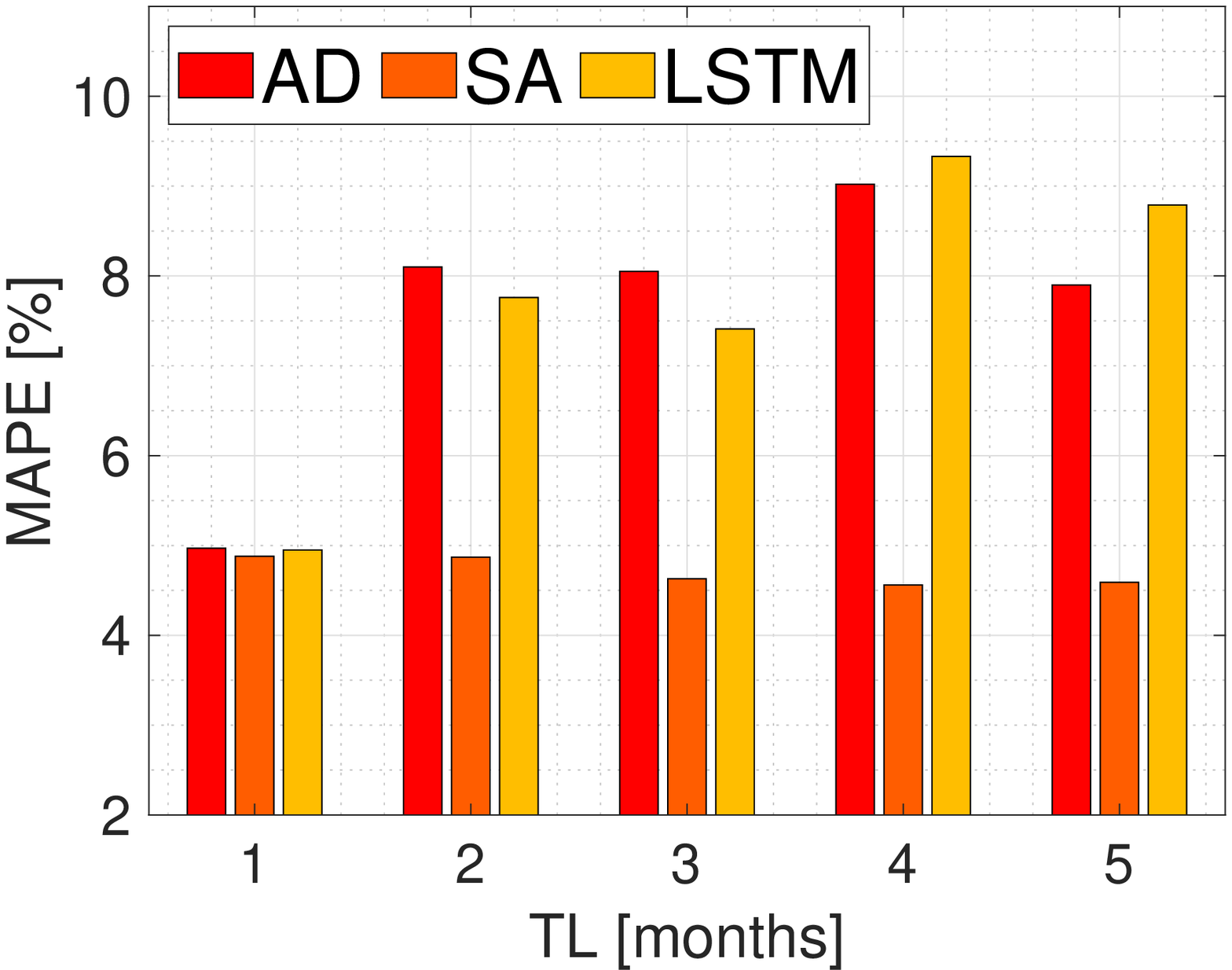}
\end{subfigure}
\caption{MAPE, CU: TL vs LA.}
\label{fig:mape_cu}
\end{figure}
\begin{figure}[t]
\begin{subfigure}
	\centering
	\includegraphics[width =.48\columnwidth]{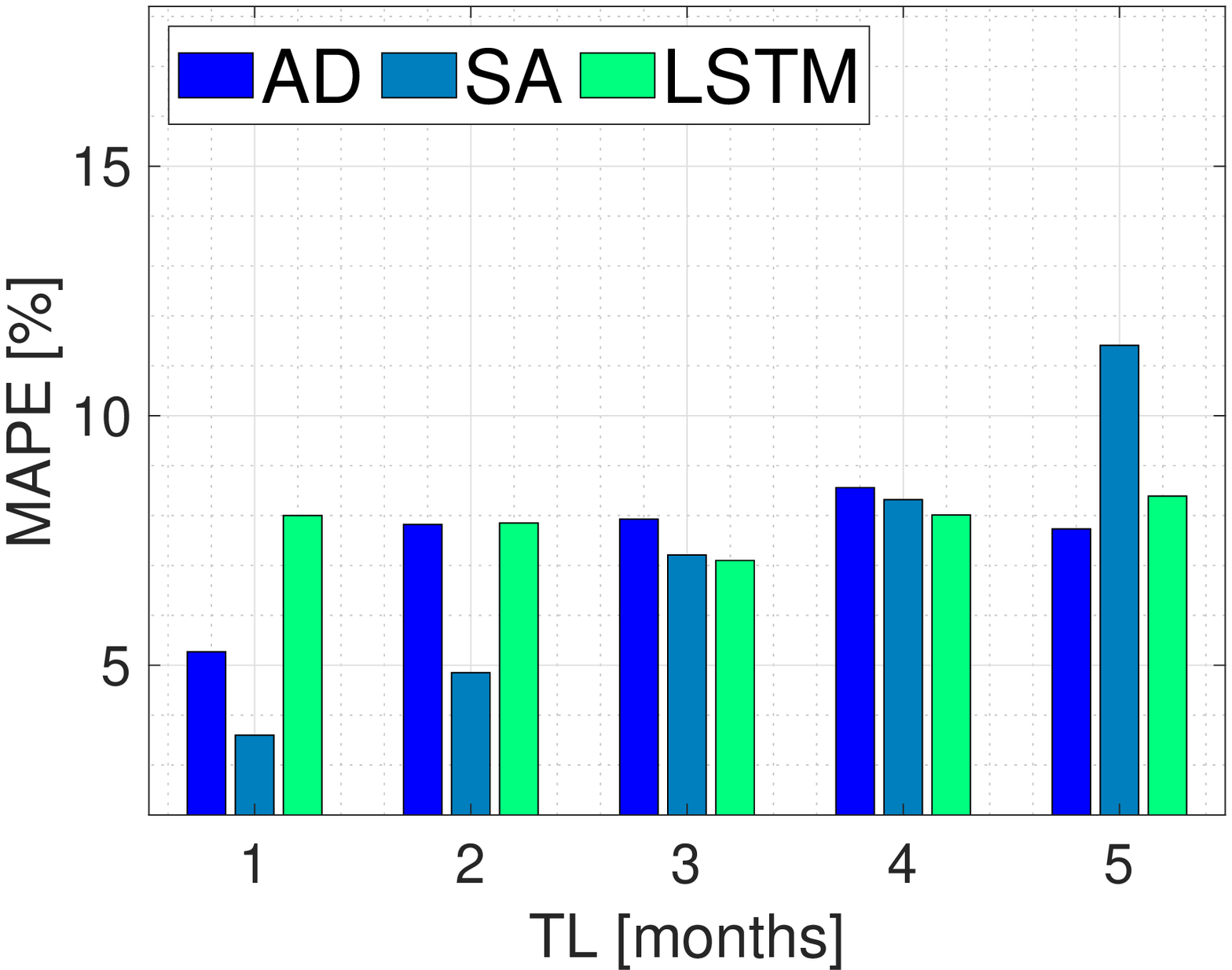}
\end{subfigure}
\begin{subfigure}
	\centering
	\includegraphics[width =.48\columnwidth]{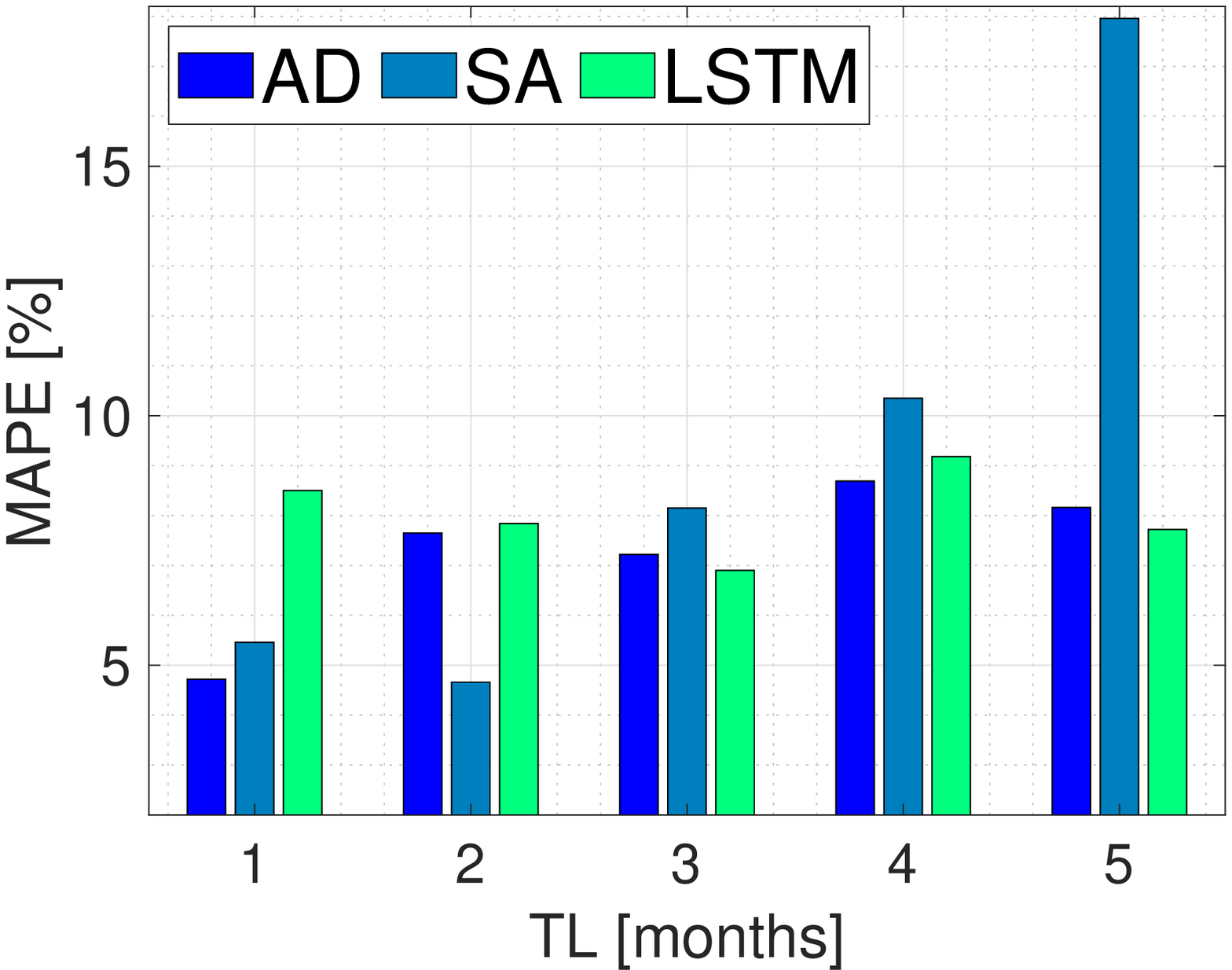}
\end{subfigure}
\caption{MAPE, CA: TL vs LA.}
\label{fig:mape_ca}
\end{figure}
As said before, to produce $\hat{\mathbf{v_{b}}}$ we consider two different approaches, Clustering Unaware (CU) and Clustering Aware (CA) respectively. According to the former approach, we train each forecasting model on $\mathbf{v_{b}}$ as defined in Section \ref{subsec:busyhour}. Differently, according to the latter approach we proceed as it follows:  
\begin{enumerate}
\item We fix the timestamps $\mathbf{t_b}$ from the vector of busy hours traffic samples $\mathbf{v_b}$; 
\item Considering the set of clusters $\mathcal{C} = [\mathcal{R}_1,\mathcal{R}_2,\mathcal{B},\mathcal{T}]$, we compute the busy-hour per-cluster cumulative downlink traffic $\mathbf{v^{c}_{b}}$ in $t^{d}_b \in \mathbf{t_b}$ for each day $d \in T\textsubscript{train}$, for $c \in \mathcal{C}$. This generates 4 different time series, namely: $\mathbf{v^{\mathcal{R}_1}_{b}}, \mathbf{v^{\mathcal{R}_2}_{b}}, \mathbf{v^{\mathcal{B}}_{b}}, \mathbf{v^{\mathcal{T}}_{b}}$;
\item We train the forecasting models on each of the four newly defined time series, independently;
\item We compute $\hat{\mathbf{v_{b}}}$ as:
\begin{align}
	& \hat{\mathbf{v_{b}}} = \sum_{c \in \mathcal{C}} \hspace{2mm} {\hat{\mathbf{v^{c}_{b}}}}.
	\end{align}
\end{enumerate}
\begin{table}[t!]
\centering
\caption{MAPEs obtained for a 1-month look-ahead and TL$\leq$2.}
\label{table:comparison_l1}
\begin{tabular}{|c|c|c|c|c|}
\hline 
\multicolumn{1}{|c}{\textbf{LA=1}} & \multicolumn{2}{|c|}{\textbf{CU}}                              & \multicolumn{2}{c|}{\textbf{CA}}  \\ \hline 
\diagbox{\textbf{Model}}{\textbf{TL}}   & \textbf{1}      & \textbf{2}     & \textbf{1}      & \textbf{2}     \\ \hline
\textbf{AD}                                           & 5.55            & 7.92           & 5.27            & 7.82           \\ \hline
\textbf{SA}                                           & 3.90             & 3.90            & 3.60             & 4.85           \\ \hline
\textbf{LSTM}                                         & 4.87            & 7.32           & 8.00              & 7.85           \\ \hline
\end{tabular}
\end{table}
\begin{table}[t!]
\centering
\caption{{MAPEs obtained for a 2-months look-ahead and TL$\leq$2.}}
\label{table:comparison_l2}
\begin{tabular}{|c|c|c|c|c|}
\hline 
\multicolumn{1}{|c}{\textbf{LA=2}} & \multicolumn{2}{|c|}{\textbf{CU}}                              & \multicolumn{2}{c|}{\textbf{CA}}  \\ \hline 
\diagbox{\textbf{Model}}{\textbf{TL}}   & \textbf{1}      & \textbf{2}     & \textbf{1}      & \textbf{2}     \\ \hline
\textbf{AD}                                           & 4.97            & 8.10           & 4.72            & 7.65           \\ \hline
\textbf{SA}                                           & 4.88            & 4.87            & 5.46             & 4.66           \\ \hline
\textbf{LSTM}                                         & 4.95            & 7.76           & 8.5               & 7.84           \\ \hline
\end{tabular}
\end{table}
We plot in Figures \ref{fig:mape_cu} and \ref{fig:mape_ca} the MAPEs obtained when either CU or CA approach is adopted, respectively. On the one hand, considering Figure \ref{fig:mape_cu}, for both LA=1 (left) and LA=2 (right) errors are below 10\% and the best performing model is SA. In particular, when SA predicts one month in advance the errors are below 5\%, with the best performance of 3.90\% obtained for a TL$\leq$2 months. Differently, when LA=2, SA performs at par for different training period lengths, with an average MAPE of 4.70\%. Performance decrease at least by 20\% when either AD or LSTM are used, with the only exception of the case (TL=1,LA=2) where their MAPEs are comparable to the one yielded by SA. On the other hand, when the knowledge about the clusters configuration is included in the forecasting process (Figure \ref{fig:mape_ca}), SA model confirms to be the best performing for both LA=1 and LA=2. In particular, best MAPE of 3.60\% is observed when SA is used with TL=1 and LA=1. Good results are also observed when TL=2 for both L=1 and L=2, where SA yields a MAPE of 4.85\% and 4.66\% respectively. Viceversa, when TL$\geq$2 SA deteriorates its performance and yields results which are comparable to the ones of the other forecasting methods (with errors above 7\%). 
To compare the performance of the two forecasting approaches, we report in Tables \ref{table:comparison_l1} and \ref{table:comparison_l2} the MAPEs obtained when TL$\leq$2 for LA=1 and LA=2, respectively. As one can see in Table \ref{table:comparison_l1}, CA approach improves by 8\% the best performance obtained when clusters are not computed. Fixing the training period, we also observe a performance improvement of 5.5\% when AD model is used, while forecasting errors increase when clusters are available if LSTM is used (regardless of TL). Similar conclusions can be drawn if we consider a look ahead of 2 months (as shown in Table \ref{table:comparison_l2}), where CA approach improves by 5\% the best performance obtained with CU approach (for TL=2). Regardless of the approach and the look-ahead capability, we observe a decreasing trend in forecasting performance with the increasing of TL, for all the tested methodologies. This is due to the particular characteristics of the period under study, as users network usage (and thus cellular traffic behaviour) has been rapidly changing during COVID-19 pandemic breakout. This suggests that in such cases it is preferable to frequently update traffic forecasting models to capture more recent variations in traffic and discard older data patterns. 
\begin{figure}[t]
\begin{subfigure}
	\centering
	\includegraphics[width =.48\columnwidth]{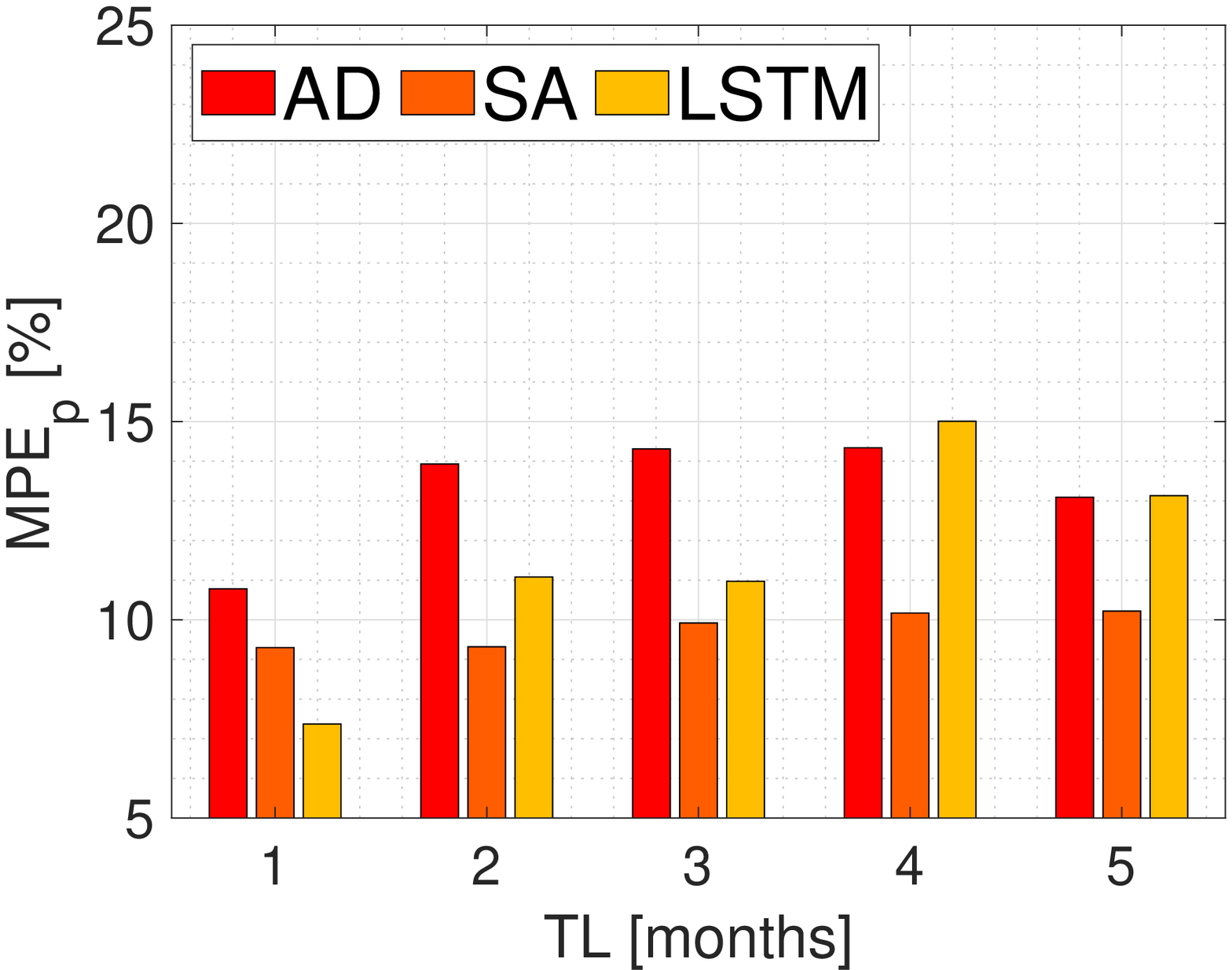}
\end{subfigure}
\begin{subfigure}
	\centering
	\includegraphics[width =.48\columnwidth]{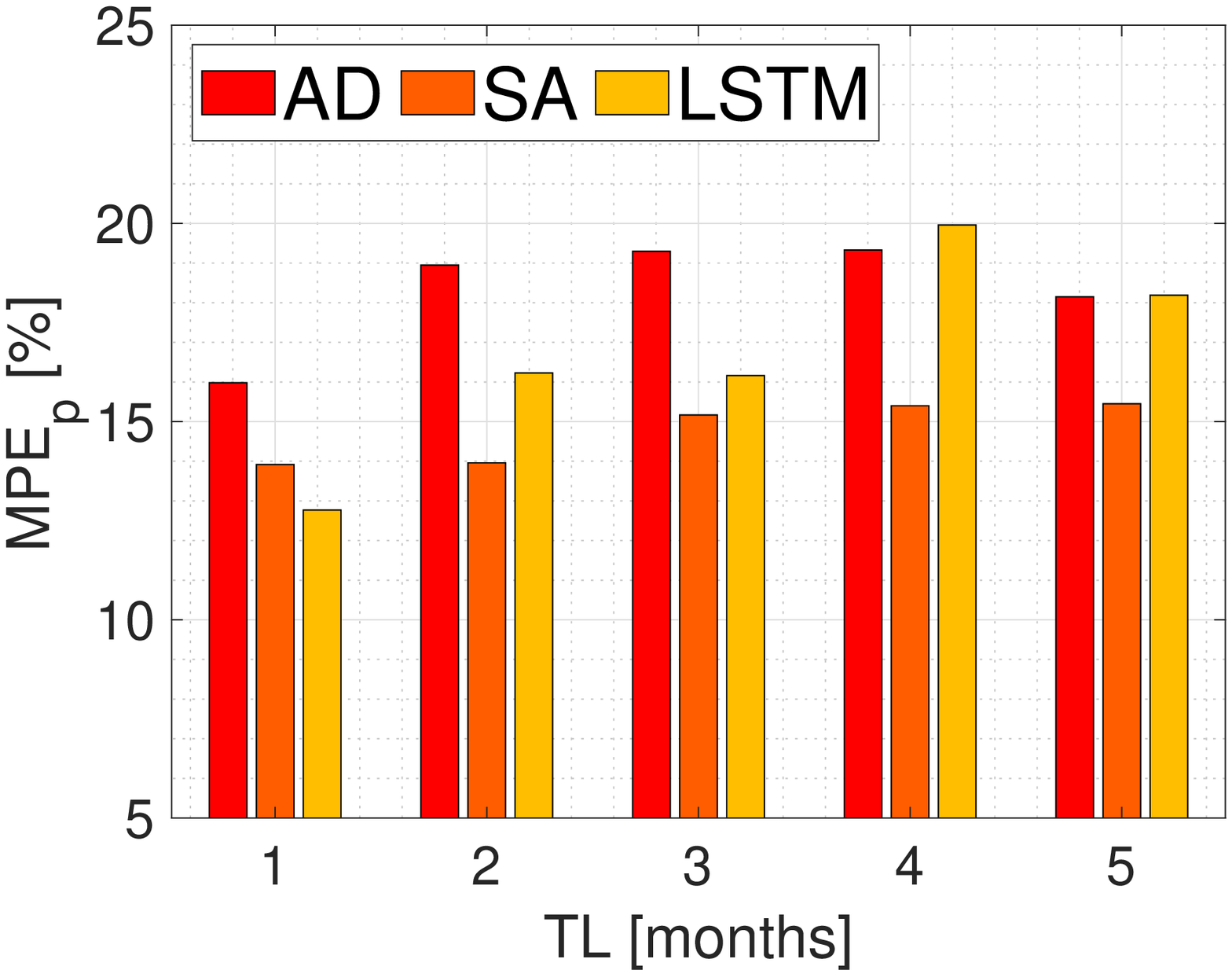}
\end{subfigure}
\caption{MPE, CU: TL vs LA.}
\label{fig:mpe_cu}
\end{figure}
To conclude, we plot in Figures \ref{fig:mpe_cu} and \ref{fig:mpe_ca} the forecasting performance with respect to the maximum busy hour downlink traffic value observed in the test period, for CU and CA approach respectively. Regardless of the forecasting approach, we observe that: i) errors are positive in sign, meaning that our models under-estimate the overall network traffic peak; ii) peak detection performance improve on average of 50\% when LA=1 with respect to LA=2. On the one hand, when LA=1, the best MPE\textsubscript{P} (7.4\%) is achieved by LSTM model for TL=1 when CU approach is adopted (11\% better than CA approach). On the other hand, when LA=2 traffic peaks are better detected (MPE\textsubscript{P}=11.88\%) when CA approach is implemented (7\% better than CU approach) and SA model is used with TL=1.
\balance
\section{Conclusions}\label{sec:conclusions}
Traffic forecasting is an important tool for MNOs to anticipate the knowledge of traffic demand peaks and thus enhance the efficiency of network dimensioning and planning operations. In this paper, we consider as target the busy hour downlink traffic and we compare the performance of different forecasting methods when either no network sites clustering is performed before training the models or a clustering configuration is designed and leveraged during the training phase. Beside the insights which the knowledge of a clustering configuration provides with respect to mobile users activity in a cellular network, our results show that it also benefits the forecasting performance, improving the accuracy by more than 5\% up to a forecasting horizon of 2 months. To conclude, we recall that the data used in this study refers to the period January-July 2020 in the country of Italy. Considering that since March the 9th Italy was in lock-down due to Covid-19 pandemic, part of the data refer to an extraordinary behaviour of the mobile network. In particular, the decreasing trend of downlink traffic is due to the reduced mobility of the users in such period: we plan to perform new experiments once the pandemic is over.


\bibliographystyle{IEEEtran}
\bibliography{IEEEabrv,bibfile}
\end{document}